\documentstyle[twoside,fleqn,espcrc2]{article}

\newcommand{\AmS}{{\protect\the\textfont2
    A\kern-.1667em\lower.5ex\hbox{M}\kern-.125emS}}
\newcommand{\Z}{{Z \!\!\! Z}}
\newcommand{\beq}{\begin{equation}}
\newcommand{\eeq}{\end{equation}}
\newcommand{\beqn}{\begin{eqnarray}}
\newcommand{\eeqn}{\end{eqnarray}}
\newcommand{\nsum}[2]{\sum_{ #1 \in \Z(\CK{#2}) }}
\newcommand{\ndsum}[2]{\sum_{\stackrel{\scriptstyle #1 \in \Z(\CK{#2})}
{\delta #1=0}}}
\newcommand{\D}{\mbox{$\cal D$}}
\newcommand{\dd}{\mbox{d}}
\newcommand{\dual}{\mbox{}^{\ast}}
\newcommand{\CK}[1]{\mbox{\scriptsize C}_{\mbox{$\scriptstyle #1$}}}
\title{Cosmic Strings on the Lattice}

\author{A. K. Bukenov\address{University of Alma--Ata,
Kazakhstan }, A. V. Pochinsky, M. I. Polikarpov\address{ITEP
Moscow, 117259, Russia},
L. Polley\address{Universit\"at
Oldenburg, 2900 Oldenburg, Germany} and
U.-J. Wiese\address{HLRZ J\"ulich, 5170 J\"ulich, Germany}}

\begin{document}

\begin{abstract}
We develop a formalism for the quantization of topologically stable
excitations in the 4-dimensional abelian lattice gauge theory. The
excitations are global and local (Abrikosov-Nielsen-Olesen) strings and
monopoles. The operators of creation and annihilation of string states are
constructed; the string Green functions are represented as a path
integral over random surfaces. Topological excitations play an important
role in the early universe. In the broken symmetry phase of the $U(1)$ spin
model, closed global cosmic strings arise, while in the Higgs phase of the
noncompact gauge-Higgs model, local cosmic strings are present. The compact
gauge-Higgs model also involves monopoles. Then the strings can
break if their ends are capped by monopoles. The topology of the Euclidean
string world sheets are studied by numerical simulations.

\end{abstract}

\maketitle

\section{INTRODUCTION}
During cooling, the early universe has undergone different phase
transitions. Depending on the symmetry that is spontaneously broken,
various excitations, for example, domain walls, cosmic strings or monopoles,
arise as topologically stable objects \cite{Vil85,Bra87}. The $U(1)$
symmetric scalar fields in the 4-dimensional field theories give rise to
string-like topological excitations. In a broken $U(1)$ symmetry phase, the
strings are stable solutions of the classical equations of motion. One
distinguishes two types of cosmic strings: global and local ones. Global
strings arise when a global $U(1)$ symmetry breaks spontaneously, whereas
local strings are due to the breakdown of the $U(1)$ gauge symmetry. A well
known example is the local Abrikosov-Nielsen-Olesen string present in the
Higgs phase of the noncompact abelian gauge-Higgs model with a local $U(1)$
symmetry \cite{Abr57,Nie73}. An even simpler example is the 4-dimensional
global $U(1)$ scalar field theory with a field $\Phi = |\Phi|
\exp(i\varphi)$. In two dimensions, zeros of the complex scalar field,
$\Phi(x) = 0$, are located at isolated points $x$. Going around $x$, the
phase $\varphi$ may change by $2 \pi n$, where $n \in \Z$ is the topological
characteristic of the vorticity of the scalar field. The relevant homotopy
group is $\Pi_1[U(1)] = \Z$. In three dimensions, the zeros of the scalar
field are located on lines (the global cosmic strings), and in four
dimensions, on surfaces (the world sheets swept out by the strings during
their time evolution). The strings are topologically stable excitations,
i.e., they are insensitive to small deformations of the field $\Phi$.

        The dynamics of cosmic strings and monopoles is usually investigated
in semiclassical physics. In the early universe, however, quantum effects
played a crucial role. The classical description is inadequate when
complicated dynamical situations arise in which strings or monopoles are
condensed. The description of phase transitions may also require the
inclusion of quantum effects. Due to the topological nature of the problem,
it is essential to formulate it in a nonperturbative framework, as
provided by the lattice regularization. For global cosmic strings, we start
with a complex scalar field $\Phi = |\Phi|^2 \exp(i \varphi)$ with a quartic
potential $V(\Phi) = \lambda (|\Phi|^2 - v^2)^2$, and consider the limit as
$\lambda \rightarrow \infty$. Then only the compact variable $\varphi \in [-
\pi,\pi]$ remains and the theory is reduced to the global $U(1)$ lattice
spin model (the four-dimensional $XY$ model). A cosmic string then manifests
itself by the vorticity of a plaquette and the corresponding dual plaquette
belongs to the Euclidean string world sheet. The local
cosmic strings results from the gauging of the $U(1)$ symmetry. On the
lattice, this leads to the noncompact and the compact abelian gauge-Higgs
models. The theory with compact gauge fields has monopoles as additional
topological excitations. Since the string theory is formulated on the
lattice, one has complete nonperturbative control of the dynamics. In
particular, one can study the string tension or the question of string
condensation in the early high temperature phase by standard lattice
techniques like numerical simulations or strong coupling expansions.

In the present publication, we only give the explicit form of the creation
operators of the strings, the details of the derivation will be published in
the subsequent paper. The quantization of the global cosmic strings is
discussed in ref.\cite{PoWi90}.

\section{GLOBAL STRINGS}

Let us consider the 4-dimensional $U(1)$ lattice spin model in the Villain
formulation \cite{Vil85}. Its partition function is given by

\beq
{\cal Z}=\nsum{l}{1}
       \int_{-\pi}^{+\pi}\D\varphi
        \exp(-\frac{\kappa}{2}\|\dd\varphi+2\pi l\|^2).  \label{ZVillain}
\eeq
We use the notations of the calculus of differential forms on the lattice
\cite{BeJo82}. $\D\varphi$ denotes the integral over all site variables,
$\varphi$; $\dd$ is the exterior differential operator, $\dd \varphi$ is the
link variable constructed as usual in terms of the site angles
$\varphi$. The scalar product is defined in the standard way, e.g., if
$\varphi$ and $\psi$ are the site variables, then
$(\varphi,\psi)=\sum_s\varphi(s)\psi(s)$, where $\sum_s$ is the sum over all
sites $s$. The norm is defined as: $\|a\|^2=(a,a)$; therefore
$\|\dd\varphi+2\pi l\|^2$ implies summation over all links.
$\nsum{l}{1}$ denotes the sum over all configurations of the integers $l$
attached to the links $C_1$. It occurs that the partition function
(\ref{ZVillain}) can be represented as follows \cite{PoWi90}:

\beq
{\cal Z}^K=\ndsum{\dual k}{2}
          \exp(-2\pi^2\kappa(\dual k, \Delta^{-1}\dual k)). \label{ZBKT}
\eeq
Where $\dual x$ denotes the object dual to $x$, the codifferential
$\delta=\dual \dd \dual$ satisfies the rule for partial integration
$(\varphi,\delta\psi)=(\dd\varphi,\psi)$. The summation in (\ref{ZBKT}) is
carried over the integer variables $k$, which belong to plaquettes $C_2$.
The condition $\delta k =0$ means that the summation is performed over the
closed two-dimensional objects defined by $k$. The surface elements interact
with each other via long range forces described by the inverse Laplacian.
Physically, these forces are due to the massless Goldstone bosons in the
broken phase of the original spin model. The action of the random surface
model is nonlocal and differs from the Nambu action which is
proportional to the surface area. It can be expected that the random surface
model is equivalent to a lattice theory of closed strings, and that the
closed surfaces are actually the string world sheets. To justify this
statement one must construct creation and annihilation operators of string
states.

\section{STRING CREATION OPERATORS}

The creation of a cosmic string (as a nonlocal object) requires the
use of nonlocal operators. Global (local) cosmic strings are surrounded by a
cloud of Goldstone (gauge) bosons, just as charged particles are surrounded
by their photon cloud. Creation operators for charged particles were first
constructed by Dirac \cite{Dir55}, whose idea was to compensate the gauge
variation of a charged field, $\Phi(x)' = \Phi(x) \exp(i \alpha(x))$, by a
contribution of the gauge field representing the photon cloud:

\beq
\Phi_c(x) = \Phi(x) \exp(i \int
d^3 y B_i(x - y) A_i(y)), \label{Dirac}
\eeq
where $\partial_i B_i(x) = \delta(x)$, and $A_i(x)' = A_i(x) + \partial_i
\alpha(x)$ is the photon field. The gauge invariant operator $\Phi_c(x)$
creates a scalar charged particle at the point $x$, together with the photon
cloud surrounding it. Our construction of string creation operators
\cite{PoWi90} is based on the same idea, and it is very similar to the
construction of soliton creation operators suggested by Fr\"ohlich and
Marchetti \cite{Fro86}. In fact, it is a generalization of their
construction of monopole sectors in 4-dimensional $U(1)$ lattice gauge
theory. We perform the duality transformation of the original theory defined
by the partition function (\ref{ZVillain}), and obtain the hypergauge
theory of integer valued fields on the dual lattice. The Wilson loop
$W({\cal C})$, constructed from the auxiliary gauge fields, is gauge
invariant, but not hypergauge invariant. We make $W({\cal C})$ hypergauge
invariant by surrounding it by the cloud of the hypergauge field. This
hypergauge invariant operator can be shown to create the closed string on
the curve $\cal C$. In terms of the variables $k$ (\ref{ZBKT}), the
expectation value of the string creation operator has the form:

\beqn
 \lefteqn{<U_{\cal C}>= \frac{1}{{\cal Z}^K} \times} \label{UCBKT}\\
\lefteqn{\sum_{\stackrel{\scriptstyle k \in \Z(\CK{2})}
{\delta \dual k=-\delta_{\cal C}} }
exp\{ -2\pi^2\kappa(\dual(k+D_{\cal C}),
\Delta^{-1}\dual(k+D_{\cal C}))\};} \nonumber
\eeqn
$D_{\cal C}$, being the counterpart of the function $B_i(x)$ in
eq.(\ref{Dirac}), depends on the curve $\cal C$: $\delta^{(3)}\dual D_{\cal
C}=\delta_{\cal C}$, since the operator of the creation of the string should
act at a definite time slice, we use the three-dimensional operator of the
codifferentiation $\delta^{(3)}$; $\delta_{\cal C}$ is the lattice delta
function which is equal to unity on the links belonging to the curve $\cal
C$, and equal to zero on the other links. It is important that because of
the condition $\delta\dual k=-\delta_{\cal C}$, the summation in
(\ref{UCBKT}) is performed over closed surfaces, and those bounded by
the curve $\cal C$. This is exactly what one would expect intuitively: a
string world sheet opens up on curve along which the string is created. In
the general case, the curve $\cal C$ may consist of several closed loops:
$\delta_{\cal C}=\sum_i\delta_{{\cal C}_i}$; then $D_{\cal C}=\sum_i
D_{{\cal C}_i}$. Moreover, if we set $\delta_{{\cal C}_j} = -1$ on several
closed loops, then the string is annihilated on these loops. Placing the
loops ${\cal C}_i$ and ${\cal C}_j$ on different time slices, we may
construct operators corresponding to string scattering and decay
processes.

        The expectation value of the string creation operator in terms of
the original fields $\varphi$ is:

\beqn \label{UCVillain} <U_{\cal C}> =
  \frac{1}{{\cal Z}}\nsum{l}{1}\int^{\pi}_{-\pi}\D\varphi \\
  \exp(-\beta\| \dd\varphi + 2\pi\delta\Delta^{-1}(D_{\cal C}-\rho_{\cal C})
   + 2\pi l \|^2). \nonumber
\eeqn
Here the integer valued field $\rho_{\cal C}$ satisfies the equation
$\delta^{(3)}(\dual D_{\cal C} - \dual\rho_{\cal C})=0 $. It can be shown
that $\dual \rho_{\cal C}$ is the analog of the (invisible) Dirac string. The
Dirac string connected to the monopole is a one-dimensional object, while
$\dual\rho_{\cal C}$, being defined on the plaquettes, is a two-dimensional
object.

\section{LOCAL COSMIC STRINGS}

        Local cosmic strings are topological excitations in the abelian
gauge-Higgs model whose partition function in the Villain form is given by:

\beqn
\lefteqn{{\cal Z}= \nsum{n}{2}\int\D\theta
        \nsum{l}{1}\int_{-\pi}^{+\pi}\D\varphi} \label{ZHVillain} \\
\lefteqn{\exp( -\beta\|\dd\theta+2\pi n\|^2
        -\frac{\kappa}{2}\|\dd\varphi-\theta+2\pi l\|^2).} \nonumber
\eeqn
If the gauge field is noncompact (the integration over $\theta$ is from
$-\infty$ to $+\infty$), then the excitations are closed local strings.
Below we discuss a more interesting case, when the gauge fields are compact
($-\pi < \theta \leq \pi$) and the monopoles are present in addition to
strings. It occurs that the strings may be open, with monopoles at their
ends. The general Green function for this case consists of the creation and
annihilation operators of strings and monopoles:

\beqn
\lefteqn{\langle \prod_i U_{{\cal C}_i}
\prod_j \Phi_{x_j} \prod_k \bar{\Phi}_{x_k} \rangle =} \label{strmon}\\
\lefteqn{\frac{1}{\cal Z} \nsum{n}{2}\int_{-\pi}^{+\pi}\D\theta
\nsum{l}{1}\int_{-\pi}^{+\pi}\D\varphi}  \nonumber\\
\lefteqn{\exp(- \beta
\|d\theta + 2 \pi n + 2 \pi \delta \Delta^{-1} (B - \omega)\|^2)}
\nonumber \\
\lefteqn{ \exp(- \frac{\kappa}{2}
\| d\varphi + 2 \pi l - \theta +
2 \pi \delta \Delta^{-1} (D - \rho) \|^2).}
\nonumber
\eeqn
It follows from the derivation of this formula (which is skipped)
that $\delta \sum_i \delta_{{\cal C}_i} + \sum_j \delta_{x_j} -
\sum_k \delta_{x_k} = 0$; therefore open strings carry monopoles and
antimonopoles at their ends. In the case of creation of monopoles at
points $x_j$ and antimonopoles at points $x_k$, the field $B$ satisfies the
equation: $\delta^{(3)}\dual B = \sum_j \delta_{x_j} - \sum_k
\delta_{x_k}$; Dirac string $\dual \omega \in \Z$ is such that $\delta^{(3)}
(\dual B - \dual\omega) = 0$; $D$ and $\rho$ are defined by the equations:
$\delta^{(3)}\dual D=\delta_{\cal C} + \dual B$; $\dd^{(3)} (D - \rho) = B -
\omega$. It is easy to show that the Green function (\ref{strmon}) is
invariant under the deformation of the Dirac string: $\omega'= \omega +
\dd\xi,\ \rho' = \rho + \xi$.

\section{COSMIC STRING DYNAMICS AND NUMERICAL SIMULATIONS}

In the high temperature phase of the early universe, the $U(1)$ symmetry was
unbroken and strings were condensed. This was possible because the string
tension vanished and string creation required no energy. As a consequence,
one expects that the strings formed clusters percolating through the
universe. After the phase transition, the $U(1)$ symmetry gets spontaneously
broken and the string network freezes. Because the string tension no longer
vanishes, the strings become massive. Small strings shrink and decay,
whereas large cosmic strings survive as stable massive structures catalyzing
the formation of galaxies. Such scenario can be studied within the formalism
discussed above.

        The string condensate can be calculated numerically by direct
measurement of the expectation value $<U_{\cal C}>$. A similar
measurement of the monopole condensate in the compact electrodynamics was
performed in refs.\cite{Pol91,PoPoWi91}. Another quantity that can be
studied numerically is the string tension of the cosmic strings. We are
performing these calculations at present.

        The simplest objects related to the cosmic strings are the defects
in the four-dimensional $XY$ model. In the given configuration of the spins
$\varphi$, each plaquette carries vorticity $\dual j =
\frac{1}{2\pi}([\varphi_1-\varphi_2]_{2\pi}+
[\varphi_2-\varphi_3]_{2\pi}+[\varphi_3-\varphi_4]_{2\pi}+
[\varphi_4-\varphi_1]_{2\pi})$, where $\varphi_k$ are the angles on the
corners of the plaquette, and $[\alpha]_{2\pi}$ is $\alpha\ \mbox{mod}\
2\pi$. If $\dual j \neq 0$, then we may ascribe vorticity $j$ to the
plaquette dual to the original one. The surfaces formed by these
dislocations are closed. The proof is very simple: $\dual j =
\frac{1}{2\pi}\dd([\dd\varphi]_{2\pi}) = \frac{1}{2\pi}\dd(\dd\varphi+2\pi
p) = \dd p$, where the integer $p$ is such that $(\dd\varphi+2\pi
p)\in(-\pi,\pi]$. Now, using equalities $\dd = \dual \delta \dual$ and
$\delta ^2 = 0$, we get $\delta j = 0$. These world sheets are characterized
by the Euler number: $N_E=n_p-n_l+n_s$, where $n_p$, $n_l$ and $n_s$ are the
number of the plaquettes, the links and the sites belonging to the world
sheet respectively. The number of handles is defined by $g=1- N_E/2$.

        We have studied numerically the topology of the world sheets in the
four-dimensional $XY$ model on the lattices $6^4 - 8^4$ for the different
values of $\kappa$. Usually, there are several disconnected closed world
sheets in each configuration of the spins. It occurs that below the phase
transition ($\kappa<\kappa_C$), in the region of the expected
condensation of the strings, there exist one world sheet of the size
comparable to that of the lattice; this world sheet contains a lot of
handles. There are also satellites: small disconnected objects with a
simple topology (no handles). After the phase transition ($\kappa >
\kappa_C$), almost all world sheets have no handles.
It turns out that the number of handles
$g_i$ taken into account with the weight proportional to the area $S_i$ of
the corresponding world sheet $i$ (i.e. $<g> = \sum_i g_i S_i/\sum_i S_i$)
is the order parameter of the system. Numerical simulations show that $<g>
\neq 0$ for $\kappa < \kappa_C$ and $<g> = 0$ for $\kappa > \kappa_C$.

Two of the authors (MIP and LP) would like to thank the HLRZ in J\"ulich for
hospitality. The work of MIP and AVP has been partially supported by the
grant of the American Physical Society.

\end{document}